\documentclass{aa}
\usepackage{txfonts,rotating}
\usepackage{graphicx}

\begin{document}

\def\msun{\hbox{M$_\odot$}}
\def\noao{\hbox{The NOAO Deep Wide Field Survey MOSAIC Data Reductions, 
http://www.noao.edu/noao/noaodeep/ReductionOpt/frames.html}}
\def\prepare{\hbox{in preparation}}
\def\inpress{\hbox{MNRAS Letters, in press, arXiv:1705.08186}}

\title{The real population of star clusters in the bar of the Large Magellanic Cloud}

\author{Andr\'es E. Piatti\inst{1,2}
}

\institute{Consejo Nacional de Investigaciones Cient\'{\i}ficas y T\'ecnicas, Av. Rivadavia 1917, C1033AAJ,Buenos Aires, Argentina;
\and
Observatorio Astron\'omico, Universidad Nacional de C\'ordoba, Laprida 854, 5000, 
C\'ordoba, Argentina\\
\email{andres@oac.unc.edu.ar}
}

\date{Received / Accepted}

\abstract{We report results on star clusters located 
in the South-Eastern half of the Large Magellanic (LMC) bar from Washington $CT_1$ 
photometry. Using appropriate kernel density estimators we detected 73 star cluster candidates, three
of which do not show any detectable trace of
star cluster sequences in their colour-magnitude diagrams (CMDs).
We did not detect other 38 previously catalogued clusters, which
could not be recognized when visually inspecting the
$C$ and $T_1$ images either; the distribution of stars in their 
respective fields do not resemble that of an stellar aggregate.
They represent $\sim$ 33 per cent of 
all catalogued objects located within the analysed LMC bar field.
From matching theoretical isochrones to the cluster CMDs cleaned from field star
contamination, we derived ages in the range
7.2 < log($t$ yr$^{-1}$) < 10.1. As far as we are aware, 
this is the first time homogeneous age estimates based on resolved stellar photometry are obtained 
for most of the studied clusters. We built the cluster frequency (CF) for the surveyed area, and 
found that the major star cluster formation activity has taken place during the period 
log($t$ yr$^{-1}$) $\sim$ 8.0 -- 9.0.  Since $\sim$ 100 Myr ago,
clusters have been formed  during few bursting formation episodes.
When comparing the observed CF to that recovered from the star formation rate
we found noticeable differences, which suggests that  field star and star
cluster formation histories could have been significantly different.}


\titlerunning{LMC bar star clusters}
\authorrunning{Andr\'es E. Piatti}

\maketitle

\markboth{Andr\'es E. Piatti: LMC bar star clusters  }{}

\section{Introduction}

Although it is expected that most of the Large Magellanic Cloud (LMC) 
clusters catalogued
by \citet[][hereafter B08]{betal08} are real extended objects, B08
did not confirmed their nature. Because they come
from inspection of photografic plates by eye or by automatic codes, we should not rule out
that some of them could be asterims.
Indeed, \citet{pb12} and \citet{p14} found 10-15\% of
catalogued objects to be possibly non-physical systems. 
Cleaning cluster catalogues os not an exciting job. Indeed, \citet{nayaketal2016}
have preferred not to study star clusters on the basis of
variation in the field star distribution or embedded in fields suffering from large
dispersion in the field star count with respect to the average,
around the cluster.
Here, we deal with star clusters located in the South-Eastern 
half of the LMC bar near the old globular cluster NGC\,1939. The region is 
one of the most densely populated by star clusters in the galaxy and most of them have not 
been studied from resolved stellar photometry so far. 

The paper is organized as follows: Section 2 describes que data set and the procedures
to obtain standardized Washington $CT_1$ photometry. We describe the search for
star clusters performed from the photometric data set in Section 3, while in Section
4 we derive cluster ages. The analysis of the derived ages is carried in Section 5,
where we introduce the intrinsic cluster formation history for the surveyed region.
Finally, Section 6 summarizes the main outcomes of this work.

\section{Observational data}

We took advantage of $CT_1$ Washington images available at the National Optical 
Astronomy Observatory  (NOAO) Science Data Management (SDM) 
Archives\footnote{http://www.noao.edu/sdm/archives.php.}, that were obtained
as part of a survey of the most metal-poor stars outside the Milky Way 
(CTIO 2008B-0296 programme, PI: Cole). The images analysed here consist of 
a 420 s $C$ and a 30 s $R$ exposures obtained with the 8K$\times$8K CCD camera
(36$\arcmin$$\times$36$\arcmin$ field) attached at the 4 m Blanco telescope 
(CTIO) under photometric conditions (seeing values are between 1.0 and 1.3, with an
average of 1.1) and at an airmass of 1.3. 

\begin{figure}
\includegraphics[width=\columnwidth]{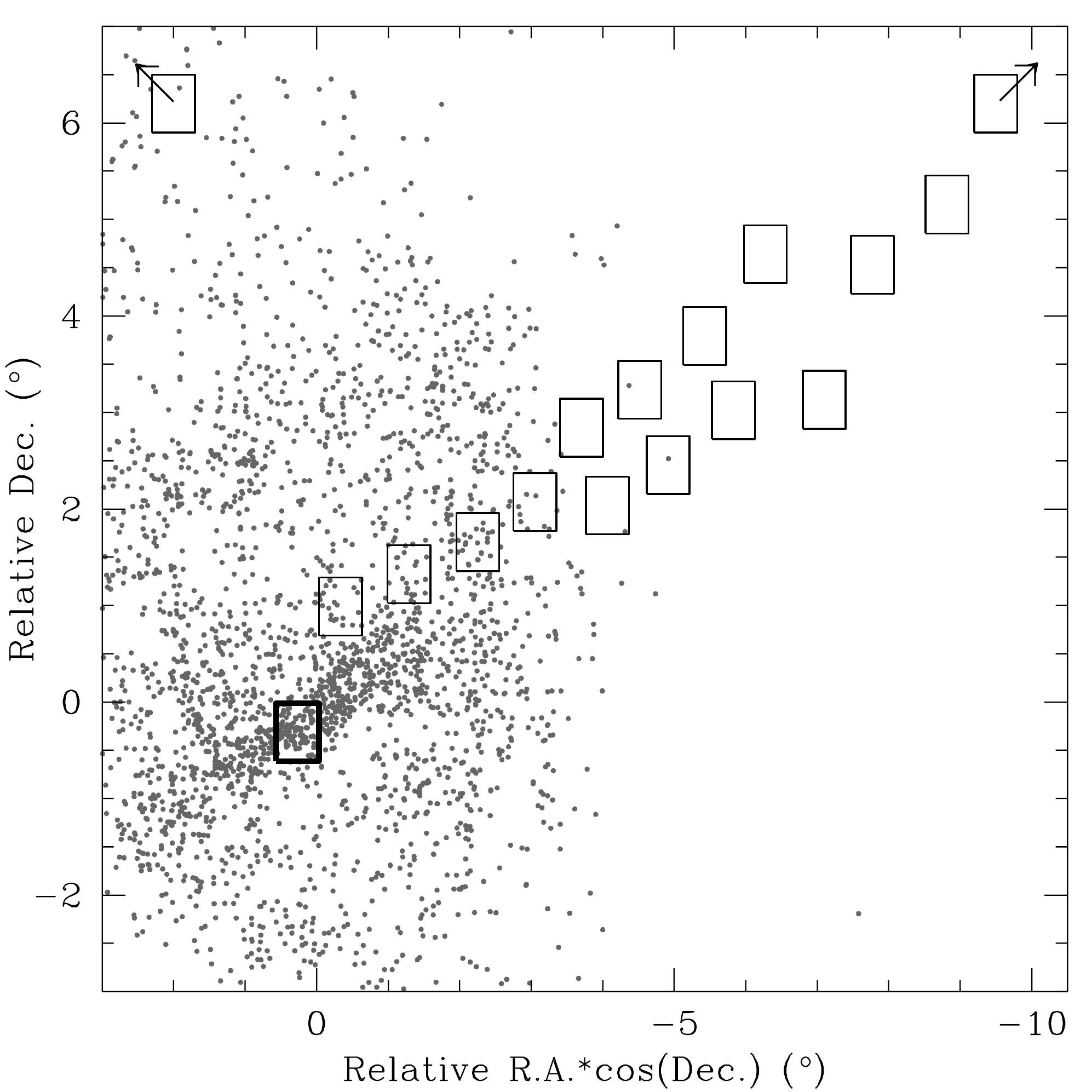}
    \caption{Spatial location of the presently studied LMC star field (thick black box), 
along with the remaining CTIO programme fields. Star clusters catalogued by \citet{betal08} are also drawn (dots) for
comparison purposes.}
   \label{fig:fig5}
\end{figure}

The data sets were fully processed following the procedures extensively described 
in \citet[e.g.][and references therein]{pietal12,p12a,p15},
together with the whole data set for the aforementioned CTIO programme, 
which comprises 17 different LMC fields (see, Fig.~\ref{fig:fig5})
and utilized the {\sc mscred} package in IRAF\footnote{IRAF is distributed by the National 
Optical Astronomy Observatories, which is operated by the Association of 
Universities for Research in Astronomy, Inc., under contract with the National 
Science Foundation.}. 
Point-spread-function photometry was obtained by employing 
the {\sc daophot/allstar}, {\sc daomatch} 
and {\sc daomaster}  suite of programs\footnote{Provided kindly by Peter Stetson.} \citep{setal90,pietal12,p15}.
The photometric errors were computed as described in \cite[e.g.][]{pb16a,pc17}.
Fig.~\ref{fig:fig2} (top-left panel) illustrates with errorbars at the
left margin typical photometric errors.  The 50 per cent completeness level is
reached at $C$ $\sim$ $T_1$ $\approx$ 20.0 mag \citep[see, e.g.][]{pc17}.

\section{Identification of star clusters}

We identified star clusters using an upgraded version of the procedure 
developed in \citet{petal2016}
and also successfully used elsewhere \citep[e.g.][]{p16,p17}. 
We particularly used a cut-off density of 
1.5 times the local background dispersion above the mean
background value.
In our case, we generated a stellar density surface over the studied region
from 902095 stars with positions and magnitudes measured in the two
$CT_1$ filters. 
Here we finally got 73 star cluster candidates, all of them
included in B08, except a new cluster candidate centred at (RA, Dec) = 
(81$\degr$.099754, -69$\degr$.609673) (J2000.0).

\begin{figure*}
\fbox{\includegraphics[width=\textwidth]{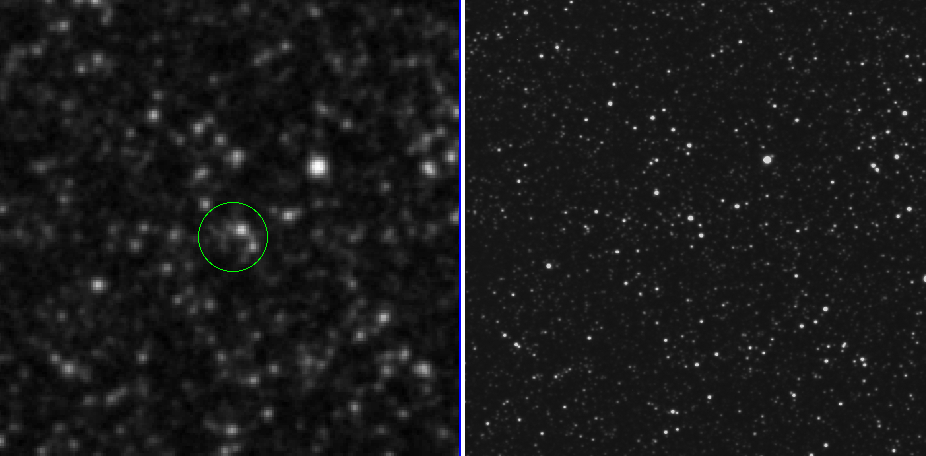}}
    \caption{3$\times$3 arcmin$^2$ DSS Red (left) and Washington $R$ (right) 
    images centred on OGLE-CL\,LMC\,414,  previously catalogued as a cluster 
and not recovered as such in the present work. North is up and East to
the left. The green circle illustrates the angular dimension given in B08.
}
   \label{fig:fig1}
\end{figure*}

We extracted from B08 every object with (RA, Dec) coordinates
within the boundaries of the surveyed area in order to check whether the 
cluster search could pass over any catalogued one.
We found that 39 catalogued clusters were not identified; one of them
([SL63] 443) because it falls on a Mosaic II image gap. The other 38 objects 
(see Table~\ref{tab:table1}) could 
not be recognized when visually inspecting the $C$ and $T_1$ images, since 
the distribution of stars in their respective fields do not resemble that of 
an stellar aggregate. We consider them as  probable non-genuine star clusters. 
The analysed crowded region shows high star field density 
variations that, in addition to the particular spatial resolution used and 
magnitude limit reached by previous cataloguing works, could lead them to 
infer the existence of extended objects (sometimes not resolved). 
Indeed, B08's catalogue includes objects discovered
by the Optical Gravitational Lens Experiment \citep[][ OGLE\,III]{u03}, whose depth
is of the order of 1.5 mag shallower than the Magellanic Cloud Photometric Survey 
 \citep[][ MCPS]{zetal04}, which in turn reaches a limiting magnitude $V$ $\sim$ 
20 mag \citep{netal09}. Our limiting magnitude is $T_1$ $\approx$ 22.5 mag
\citep{petal2017}. As an example, Fig.~\ref{fig:fig1} 
compares an enlargement of the $R$ image centred on OGLE-CL\,LMC\,414 to that 
obtained from the DSS Red one. The version of the figure with all objects listed in
Table~\ref{tab:table1} is available as Supporting Information online.
The 38  probable non-genuine physical systems represent $\sim$ 33 per cent of all objects 
located within the analysed LMC bar field, catalogued by B08.
This percentage is much higher than those found by \citet{pb12} and \citet{p14}
for other Magellanic Clouds regions.

\begin{table}
\caption{ Probable non-genuine objects in the B08's catalogue.}
\label{tab:table1}
\begin{tabular}{@{}ccc}\hline\hline
Cluster name & Cluster name & Cluster name \\
\hline
BSDL\,1340 &[HS66]\,252        &OGLE-CL\,LMC\,434\\
BSDL\,1353 & [HS66]\,255       &OGLE-CL\,LMC\,435\\
BSDL\,1522 &[HS66]\,259        &OGLE-CL\,LMC\,437\\
BSDL\,1540 &OGLE-CL\,LMC\,375   &OGLE-CL\,LMC\,439\\
BSDL\,1592 & OGLE-CL\,LMC\,406  &  OGLE-CL\,LMC\,441\\
BSDL\,1597 &OGLE-CL\,LMC\,410   & OGLE-CL\,LMC\,443 \\
BSDL\,1614 &OGLE-CL\,LMC\,412   &OGLE-CL\,LMC\,448\\
BSDL\,1636 &OGLE-CL\,LMC\,414   &OGLE-CL\,LMC\,455\\
BSDL\,1647 &OGLE-CL\,LMC\,421   &OGLE-CL\,LMC\,465\\
BSDL\,1680 &OGLE-CL\,LMC\,425   &OGLE-CL\,LMC\,466\\
BSDL\,1681 &OGLE-CL\,LMC\,428   &OGLE-CL\,LMC\,474\\
BSDL\,1768 &OGLE-CL\,LMC\,430   &  OGLE-CL\,LMC\,475\\  
BSDL\,1784 &OGLE-CL\,LMC\,433   &  \\\hline
\end{tabular}
\end{table}

\section{Star cluster CMD cleaning}


We statistically suctracted field stars from the cluster CMDs by applying the
procedure 
developed by \citet{pb12}, and successfully used elsewhere 
\citep[e.g.][and references therein]{p14,petal15a,petal15b,pb16a}.
Here we used four star-field CMDs constructed from stars 
 within circles placed to the North, East, South and West, adjacent to the cluster region, 
 and with areas equal to the circular area (typically
with radii 2-3 times the cluster radius) used for
the cluster region. 
As a result, three objects (BSDL\,1719, [HS66]\,250 and [HS66]\,291) -whose
cleaned CMDs do not show any detectable trace of star cluster sequences- 
were discarded.

\begin{figure*}
	\includegraphics[width=\textwidth]{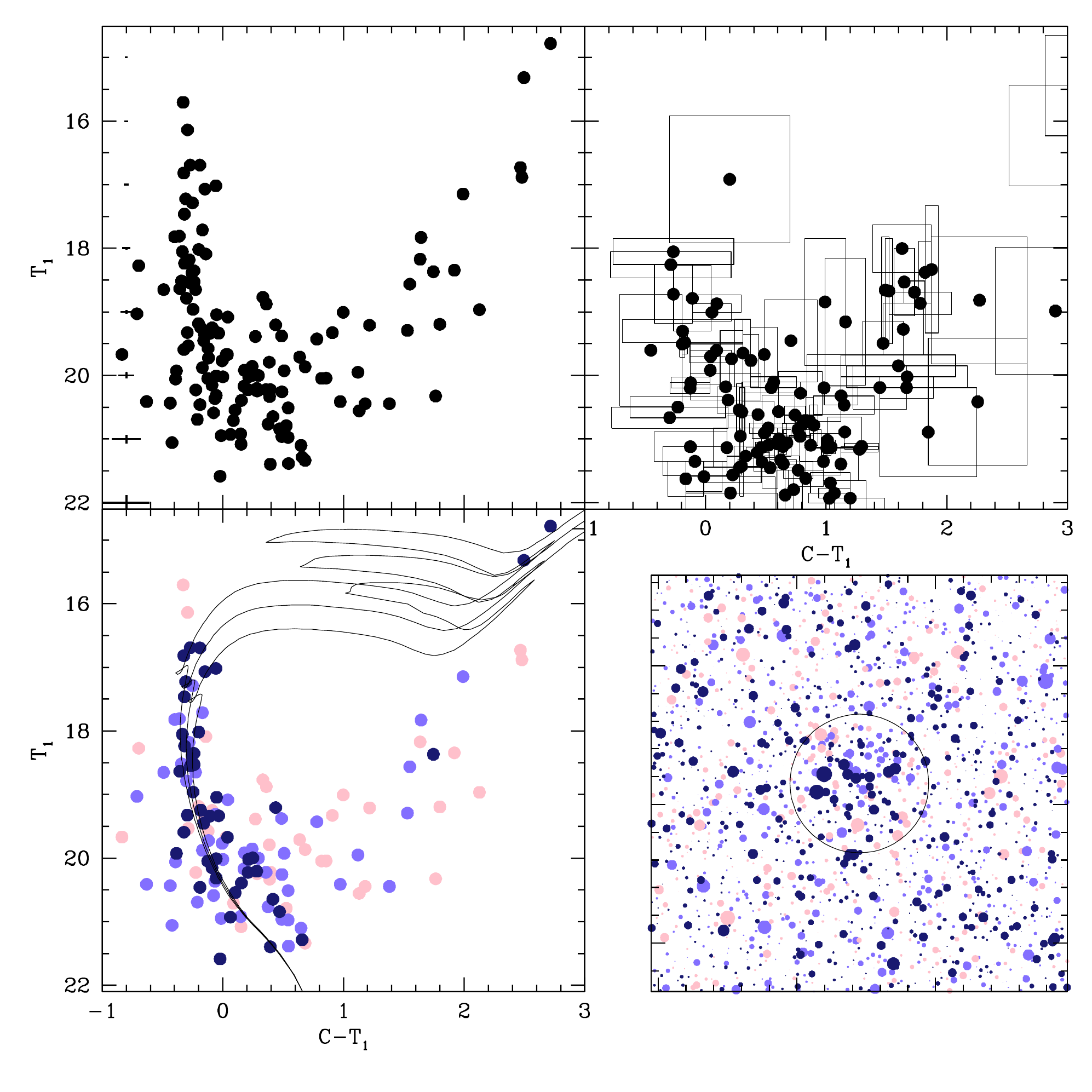}
    \caption{CMDs for stars in the field of OGLE-CL\,LMC\,377: the observed CMD
  composed of the stars distributed within the cluster radius,  with typical
 photometric errors represented with errorbars at
the left margin (top
  left-hand panel); a field CMD for a circular area equal to that of the cluster
with the respective sample of produced boxes used in the cleaning procedure
  (top right-hand panel); the cleaned cluster CMD (bottom
  left). Colour-scaled symbols represent stars with membership probability of
$P \le$ 25\% (pink), $P =$ 50\% (light blue) and $P \ge$ 75\% (dark
  blue). 
  Three isochrones from \citet{betal12} for log($t$ yr$^{-1}$)
  = 8.1, 8.2, and 8.3 and $Z$ = 0.006 are also superimposed.  
  The schematic diagram centred on the cluster is
  shown in the bottom right-hand panel. 
  The black circle represents
  the adopted cluster radius. 
  Symbols are as in the bottom left-hand
  panel, with sizes proportional to the stellar brightnesses. North is
  up; East is to the left.  The actual images are shown in Fig. A.1.
    }
   \label{fig:fig2}
\end{figure*}

Figure~\ref{fig:fig2} illustrates the performance of the cleaning
procedure for OGLE-CL\,LMC\,377.
The 70
individual photometric catalogues for the confirmed clusters are
provided in the online version of the journal. The columns of each
catalogue successively lists the star ID, the R.A. and Dec., the magnitude
and error in $C$ and $T_1$, respectively, and the photometric membership
probability ($P$). The latter is encoded with numbers 1, 2, 3 and 4 to
represent probabilities of 25, 50, 75 and 100 per cent, respectively.

According to \citet[][see their figure 6]{pg13}, LMC star clusters mostly 
expand the age range log($t$ yr$^{-1}$) $\la$ 9.40, with the exception of 
ESO\,121-SC-03 (log($t$ yr$^{-1}$) $\sim$ 9.92) and 15 old globular clusters
(log($t$ yr$^{-1}$) $\sim$ 10.1).
Young star clusters are distinguished in the CMDs by their bright MSs, while
intermediate-age clusters (9$<$log($t$ yr$^{-1}$)< 9.45) have
MS turnoffs (TOs) that decrease in brightness as they become older. A typical 
2.5 Gyr old LMC cluster (log($t$ yr$^{-1}$)=9.4) has its MS TO at $T_1$ $\sim$
20.5 mag. By assuming a depth of thet LMC disc of
(3.44$\pm$1.16) kpc \citep{ss09} and that such a cluster 
were located behind the LMC, its MS TO would result $\Delta$$T_1$ $\la$ 0.3 mag 
fainter. This means that the faintest cluster MS TO stars typically seen in the LMC
are brighter than $T_1$ $\approx$ 21.0 mag. This magnitude is even brighter 
that our limiting magnitude, so that we were able to
detect any star cluster (with stars from its  brightest limit down to its MS TO)
located in the surveyed field.

\section{Star cluster ages}

We estimated the ages of the confirmed star clusters using their CMDs
built from stars with membership probabilities higher than 50 per cent and
matching them with the theoretical isochrones of \citet{betal12}.
In performing this task, we dealt with their reddenings, distances
and metallicities. 
As for the cluster metallicies, we adopted a value of [Fe/H] = -0.4 dex 
for all of them
\citep{pg13}. 
Consequently, should we allow the metallicity to vary, 
we would not be able to see any meaningful difference along the cluster MSs, 
because of the dispersion of the stars. We made one exception in the employment
of isochrones for the old globular cluster NGC\,1939, for which we adopted
[Fe/H] = -2.0 dex.

We took advantage of the Magellanic Clouds (MCs) extinction values based
on red clump (RC) and  RR Lyrae stellar photometry provided by the OGLE\,III 
collaboration, as described in \citet{hetal11}, to estimate $E(V-I)$ colour 
excesses. We recall that they found very low reddenings in the LMC bar region.
In matching isochrones, we started by adopting those $E(V-I)$ values,
combined with the equations 
$E(V-I)$/$E(B-V)$ = 1.25, $A_{V}$/$E(B-V)$ = 3.1 \citep{cetal89}; 
$E(C-T_1)$/$E(B-V)$ = 1.97 and $A_{T_1}$/$E(B-V)$ = 2.62 \citep{g96}.
Note that considering the LMC disc depth, the difference in distance
modulus could be as large as $\Delta$$(m-M)_o$ $\sim$ 0.3 mag, which is
of the order of the uncertainties while adjusting isochrones to the
cluster CMDs in magnitude
(nearly twice as big as the size of the plotting 
symbols in Fig.~\ref{fig:fig2}), so that our simple assumption of adopting 
the same distance of all clusters should not affect the results.

Table~\ref{tab:table2} lists the derived $E(V-I)$
colour excesses and ages, while Fig.~\ref{fig:fig2} (bottom-left panel) illustrates
the performance of the isochrone matching. We estimated an upper value for our age
uncertainties of $\Delta$log($t$ yr$^{-1}$) = $\pm$0.10.

\begin{table*}
\caption{Fundamental properties of the star cluster sample. }
\label{tab:table2}
\begin{tabular}{@{}lcclcc}\hline\hline
Cluster name &  $E(V-I)$$^a$ & log($t$ yr$^{-1}$) & Cluster name &  $E(V-I)$$^a$ & log($t$ yr$^{-1}$)  \\
\hline
BRHT\,50a          &0.04   &  8.15 &NGC\,1959          &0.04   &   8.70 \\
BSDL\,1291         &0.04   &   8.20  &     NGC\,1958          &0.05   & 8.60  \\  
BSDL\,1299         &0.04   &  8.40   &NGC\,1969          & 0.04  &   8.30 \\
BSDL\,1335         &0.11   & 8.00   &NGC\,1971          &0.03   &  8.20  \\
BSDL\,1367         &0.03   &  8.40  &NGC\,1972          &0.03   &  8.40  \\
BSDL\,1381         &0.03   &    8.20 & OGLE-CL\,LMC\,369  &0.04   & 8.55 \\
BSDL\,1480         &(0.20) & 7.30    & OGLE-CL\,LMC\,376  &(0.20) & 7.30  \\
BSDL\,1491         &(0.15) &  8.30  &OGLE-CL\,LMC\,377  &0.08   &  8.20 \\
BSDL\,1511         &0.07   &  8.20  &OGLE-CL\,LMC\,396  & 0.08  &   8.55\\
BSDL\,1516         &0.07   &  8.15 &OGLE-CL\,LMC\,398  &(0.30) &  8.20  \\
BSDL\,1576         & 0.04  &   8.60 & OGLE-CL\,LMC\,400  &(0.25) &  8.00 \\
BSDL\,1601         &0.05   &  8.20  &OGLE-CL\,LMC\,402  &(0.20) & 8.15\\
BSDL\,1608         &0.04   &  8.10   &OGLE-CL\,LMC\,403  &(0.20)  &  8.25 \\
BSDL\,1707         &0.06   & 9.00   &OGLE-CL\,LMC\,407  &0.04   &  8.70 \\
BSDL\,1712         &(0.10)  & 8.00   &OGLE-CL\,LMC\,415  &(0.15) & 8.15 \\
BSDL\,1723         &0.04   & 8.35  &OGLE-CL\,LMC\,416  &  0.09 &     8.20  \\
BSDL\,1772         &0.04   &   8.50  &OGLE-CL\,LMC\,418  &(0.15) & 8.55 \\
BSDL\,1778         &0.03   &  8.75 &OGLE-CL\,LMC\,419  & 0.12  & 8.05 \\
BSDL\,1785         &0.04   &  8.35 & OGLE-CL\,LMC\,420  &0.04   &    8.70\\
H88\,283           &0.03   &  8.55 &OGLE-CL\,LMC\,429  & 0.02  &   8.40 \\
H88\,295           &0.02   & 8.75  &OGLE-CL\,LMC\,431  &0.04   &   8.05\\
\rm [HS66]\,251         &0.02   &  8.55 &OGLE-CL\,LMC\,438  &(0.10)  &8.65 \\
KMK88\,48         &0.04   &8.90    &OGLE-CL\,LMC\,442  &0.04  &  9.00 \\
KMK88\,49          &0.09   &  8.70  &OGLE-CL\,LMC\,447  &0.02 & 8.40 \\
KMK88\,50          &0.09   &  8.75 &  OGLE-CL\,LMC\,451  &0.07   &  8.80 \\
KMK88\,51          &0.12   & 8.30   & OGLE-CL\,LMC\,456  &0.04   &   8.60\\
KMK88\,52          &(0.15) &8.05   &OGLE-CL\,LMC\,462  &0.05 &   8.70  \\
KMK88\,55          &0.08   & 8.20   &OGLE-CL\,LMC\,463  &0.06   &  8.60 \\
KMK88\,56          &(0.15) & 8.45  &OGLE-CL\,LMC\,467  &0.05   &   8.25 \\
KMK88\,57          &(0.20) &8.55   &OGLE-CL\,LMC\,468  &0.06   &  8.20 \\
newcls             & 0.04  &   8.10 & OGLE-CL\,LMC\,469  & 0.07  &   8.70 \\
NGC\,1926          &0.03   &  8.35 &OGLE-CL\,LMC\,472  &0.05   &  7.60  \\
NGC\,1938          &0.07   &   8.70 &OGLE-CL\,LMC\,478  &0.05  &   8.65\\
NGC\,1939          &0.07   &  10.10 &OGLE-CL\,LMC\,479  &0.06  &  8.20 \\
NGC\,1950          &0.04   &   8.70 &[SL63]\,436        &0.04   & 8.60 \\
 \hline
\end{tabular}

\noindent $^a$ $E(V-I)$ values in parentheses
are slightly larger than those from \citet{hetal11} to get a better isochrone
matching. Nevertheless they are within the dispersion given for the OGLE\,III
$E(V-I)$ colour excesses.

\end{table*}

\section{Star cluster analysis}

Few clusters in our sample have previously been studied from resolved 
stellar photometry. \citet{mg04} presented $HST$ data which resulted
in high accuracy CMDs for NGC\,1938 and NGC\,1939. Our $CT_1$ photometry 
confirms their
results for the old globular cluster NGC\,1939 (log($t$ yr$^{-1}$) = 10.1, 
[Fe/H] =
-2.0 dex) and gives an age slightly older and within the quoted uncertainties than the
value derived by them (log($t$ yr$^{-1}$) $\sim$ 8.6) for
NGC\,1938. \citet{dg2000} obtained 
Gunn $g,i$ photometry at the ESO/MPI 2.2 m telescope (La Silla) for the
triple system NGC\,1969, 1971 and 1972, and derived ages of log($t$ yr$^{-1}$) =
7.8, 7.8 and 7.6 
with a typical error of $\sigma$(log($t$ yr$^{-1}$)) = $\pm$0.1, respectively, from the matching of 
theoretical isochrones. These values are younger than those derived here, and
could be mostly affected by star field contamination; particularly
of bright field stars assumed to be cluster stars (see their figure 7). Note
that they did not perform any decontamination of field stars in their CMDs.

\begin{figure}
	\includegraphics[width=\columnwidth]{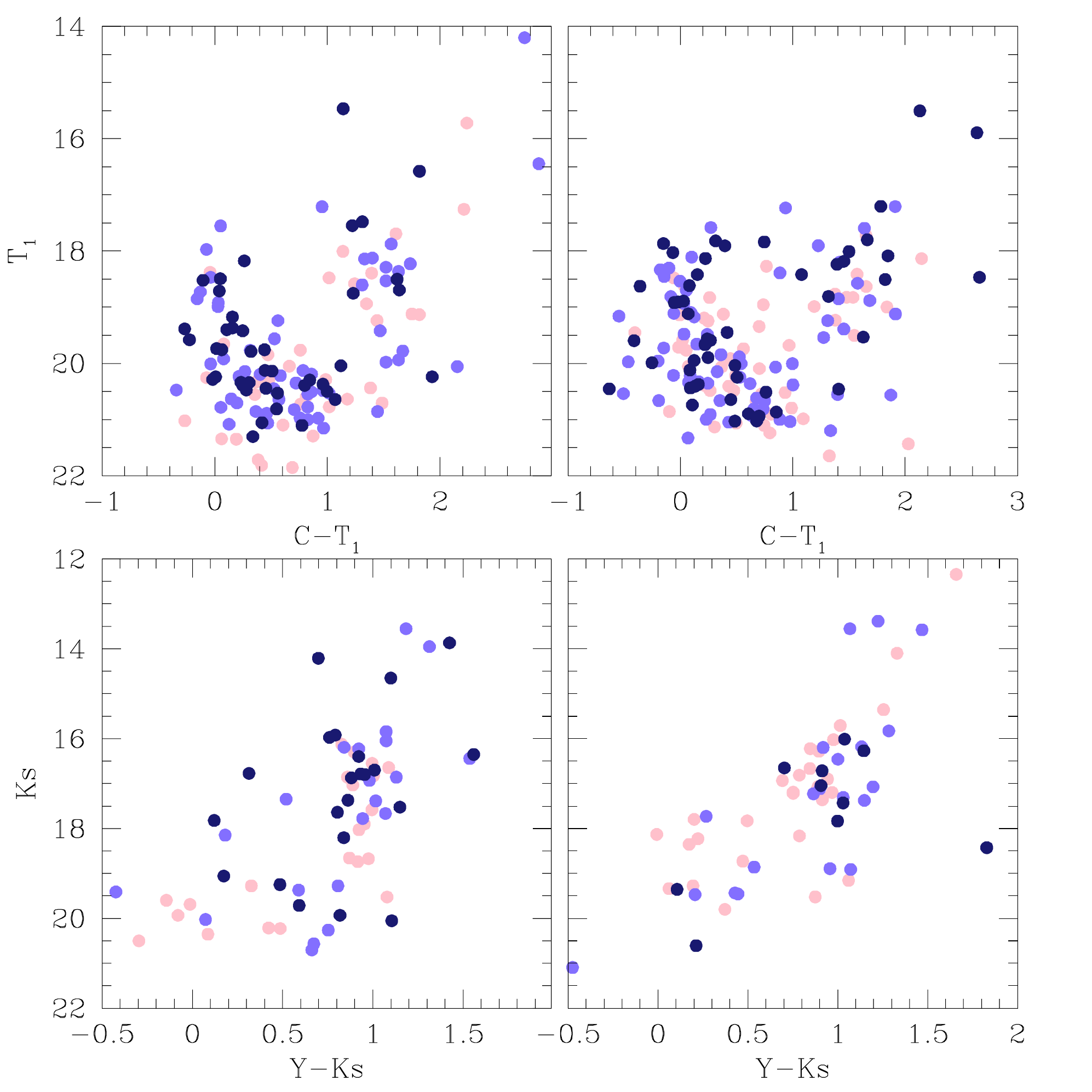}
    \caption{Cleaned CMDs for OGLE-CL\,LMC\,451 (left) and OGLE-CL\,LMC\,469 (right)
in the Washington $CT_1$ (top) and $YK_s$ (bottom) filters.}
   \label{fig:fig3}
\end{figure}

The VISTA\footnote{Visible and Infrared Survey Telescope for Astronomy.} 
near-infrared $YJK_s$ survey of the MCs system \citep[][ VMC]{cetal11} has also
imaged three clusters of our sample, namely: KMK88\,55, OGLE-CL\,LMC\,451 
($\equiv$ [HS66]\,282) and
OGLE-CL\,LMC\,469 ($\equiv$ [HS66]\,295). They were studied by \citet{petal14b} 
from CMDs built 
using PSF photometry on homogenised deep tile images \citep{retal12}.
KMK88\,55 turned out to be a cluster of  log($t$ yr$^{-1}$) $\sim$ 8.5, while OGLE-CL\,LMC\,451 and
OGLE-CL\,LMC\,469 were classified as  probable non-genuine star clusters. The older
age derived for KMK88\,55 is affected by the lack of measurements of blue cluster
stars, while the assessment on the physical reality of OGLE-CL\,LMC\,451 and
OGLE-CL\,LMC\,469 is based on a shallower VMC $K_s$ limiting magnitude. 
We show in Fig.~\ref{fig:fig3} the cleaned CMDs constructed by \citet{petal14b} 
compared to ours. 

Most of the remaining clusters in our sample, as well as those  probable non-genuine clusters of 
Table~\ref{tab:table1}, do have only age estimates on the basis
of integrated colours \citep{pandeyetal2010,poetal12}. However, \citet{asad13} showed that
unresolved methods (integrated, broad-band colour photometry) poorly match the ages
of LMC clusters derived from resolved stellar photometry (CMD). 
\citet{p14c} also found results similar to those of \citet{asad13} when
integrated spectroscopy is used to estimate cluster ages.
 
The star cluster frequency (CF) - the number of clusters per time
unit as a function of age - is a straightforward way to compare the cluster formation
activity in different epochs of the galaxy lifetime. In the case of the LMC, it has been
built for different regions and resulted to vary from one place to another
\citep[][ and references therein]{p14,p14b}. Moreover, variations within the LMC bar
has also been found \citep[e.g.][]{petal15b}. Therefore, aiming at tracing the intrinsic
cluster formation history in the surveyed area, we built its CF from the ages estimated for
the 70 studied star clusters.

\begin{figure}
	\includegraphics[width=\columnwidth]{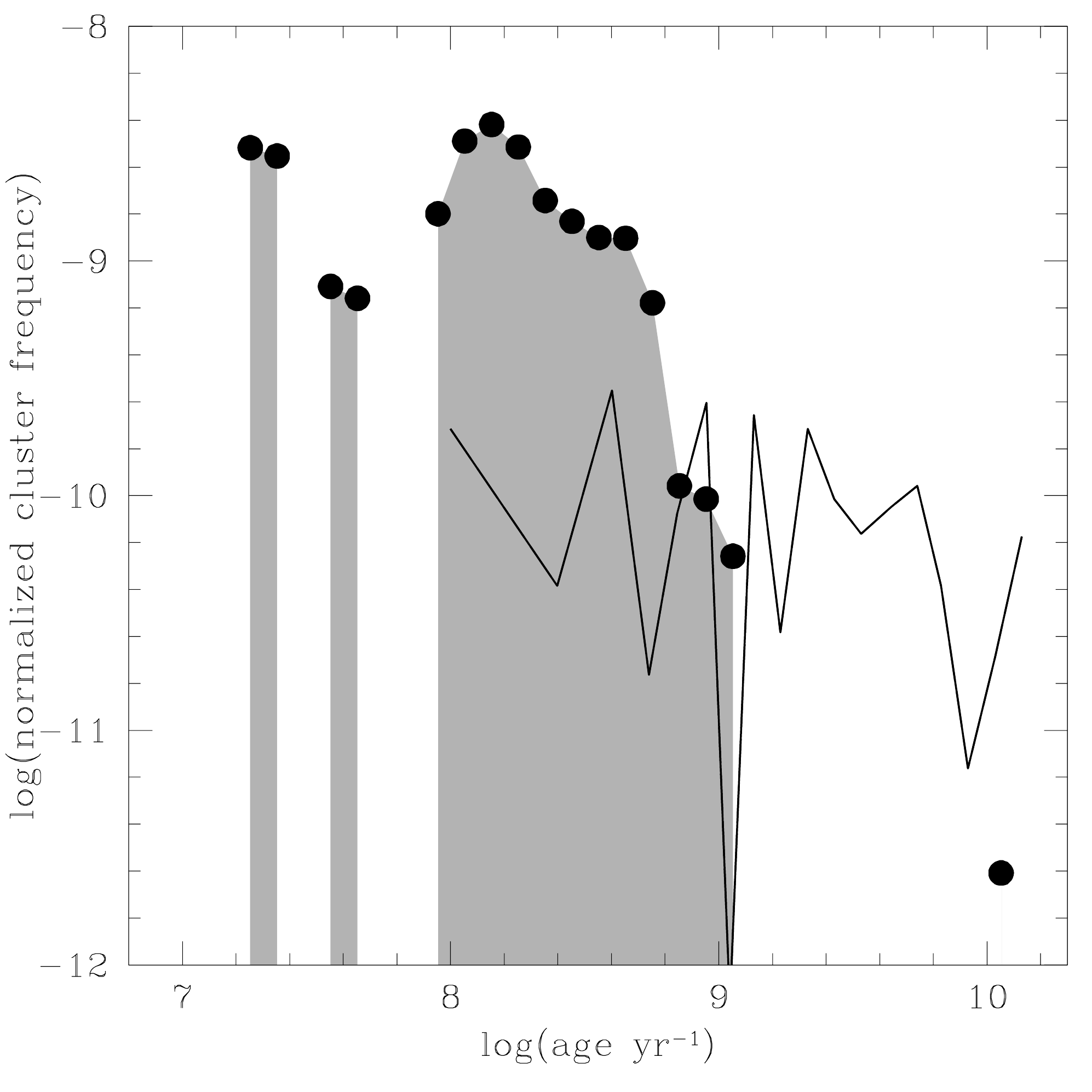}
    \caption{CF of the surveyed region in the LMC bar (filled circles). The grey areas
highlight the periods of star cluster formation activity, while the solid line
represents the CF recovered from the corresponding SFR obtained by \citet{smeckeretal2002}.}
   \label{fig:fig4}
\end{figure}

Instead of constructing an age histogram 
we assigned to each cluster a
{\it Gaussian} distribution centred on the mean cluster age and with FWHM twice as big as 
the age uncertainty. 
The result of summing the contribution of all {\it Gaussian} distributions
is depicted in Fig.~\ref{fig:fig4}. For comparison purposes, the CF was normalized to the total number of clusters.
As can be seen, the major star cluster formation activity has taken 
place during the period log($t$ yr$^{-1}$) $\sim$ 8.0 -- 9.0,
 suggesting 
that either clusters in this bar region have been formed relatively recently, or  any 
cluster older than log($t$ yr$^{-1}$) $\sim$ 9 has been disrupted. The only exception is 
the old globular cluster NGC\,1939,
which could also be an outer disc cluster projected on the LMC bar \citep{shetal10}.
Nevertheless, since \citet{p14b} found for the whole LMC bar
that there has been cluster formation activity from
log($t$ yr$^{-1}$) $\sim$ 9.4,
we conclude that this part of the bar is in average a relatively particular younger one.
The cluster formation during the last $\sim$ 100 Myr shows some short isolated 
periods of true activity. 

The derived CF was finally compared
with that obtained
from the star formation rate (SFR) derived by \citet{smeckeretal2002} 
using $HST$ observations. We used their SFR and cluster masses from 
 $\log(M_{\rm cl} [{\rm M}_\odot]) = 2.2$ to
$\log(M_{\rm cl} [{\rm M}_\odot]) = 5.0$ \citep{dgetal08,getal11}.
Fig.~\ref{fig:fig4} shows the resulting, recovered CF drawn with a solid line. 
The observed and recovered CFs are clearly different for a couple of reasons.
On the one hand, the recovered CF shows star formation activity where
there is no cluster (log($t$ yr$^{-1}$) $\ga$ 9.0). At a first glance, it 
could be somehow surprising, taking into account the common notion that most
of the stars more massive than 0.5 0.5 M$_\odot$ may form in
  clusters,
so that a significant fraction of 
  field stellar populations originate from disrupted clusters \citep[e.g.][]{ll03}. 
However, the LMC exhibits a
well-known gap in the cluster age distribution between
log($t$ yr$^{-1}$) $\sim$ 9.5 -- 10.1, while the age distribution of the field stellar
population appears more continuous \citep{pg13}. Furthermore, 
numerous authors have asserted
that the LMC's field star and star cluster formation histories are
significantly different \citep[e.g.][and references
  therein]{oetal96,getal98,s98}. 

On the other hand, the observed CF shows a noticeable excess respect to the
recovered one for ages younger than log($t$ yr$^{-1}$) $\sim$ 9.0. Even though 
the recovered CF requires additional
refinements, the observed
disparities between the cluster and field star age distributions
seem to offer evidence in support of a decoupling between star cluster 
and field star formation.
These results provide some
clues for a better approach in the study of the field stars origin and
its link to cluster disruption and environmental conditions.

\section{Conclusions}

In this work we analyse CMDs of star clusters located 
in the South-Eastern half of the LMC bar from a Washington $CT_1$ 
photometric data set. 

We performed a procedure for the star cluster search which consists 
in using {\it Gaussian} and
{\it tophat} KDEs with a bandwidth of 0.4 arcmin, and detected  
73 star cluster candidates.
We did not detect other 38 previously catalogued clusters, which
could not be recognized when visually inspecting the
$C$ and $T_1$ images either. The distribution of stars in their 
respective fields do not resemble that of an stellar aggregate.
We consider them as probable non-genuine star clusters.
The 38  probable non-genuine physical systems represent $\sim$ 33 per cent of 
all catalogued objects located within the analysed LMC bar field.

The CMDs of the star cluster candidates were statistically cleaned 
from field star contamination.
Three objects, whose CMDs do not show any detectable trace of
star cluster sequences, were discarded. The confirmed clusters comprises a
complete sample, since we were able to detect any star cluster with stars from its
brightest limit down to its MS TO located in the surveyed field.
From matching theoretical isochrones to the cleaned cluster CMDs we estimated ages
taking into account the LMC mean distance modulus, the present day metallicity and the individual star
cluster colour excesses. As far as we are aware, these are the first age
estimates based on resolved stellar photometry for most of the studied 70
clusters. The derived ages are in the age range 7.2
< log($t$ yr$^{-1}$) < 9.1, in addition to the old globular cluster NGC\,1939. 

Finally, we built the CF aiming at tracing the
intrinsic cluster formation history of the surveyed area. 
We found that
the major star cluster formation activity has taken place during the period 
log($t$ yr$^{-1}$) $\sim$ 8.0 -- 9.0, which results in average relatively
younger than the whole formation period of the LMC bar. Since $\sim$ 100 Myr ago,
clusters have been formed  during few bursting formation events.
When comparing the observed CF to that recovered from the SFR derived
by \citet{smeckeretal2002} we found noticeable differences. We conclude that
they are evidence of  field star and star
cluster formation histories are significantly different.

\begin{acknowledgements}
 We thank the referee for his thorough reading of the manuscript and
timely suggestions to improve it.
\end{acknowledgements}

\bibliographystyle{aa}

\begin{thebibliography}{44}
\expandafter\ifx\csname natexlab\endcsname\relax\def\natexlab#1{#1}\fi

\bibitem[{{Asa'd} {et~al.}(2013){Asa'd}, {Hanson}, \& {Ahumada}}]{asad13}
{Asa'd}, R.~S., {Hanson}, M.~M., \& {Ahumada}, A.~V. 2013, \pasp, 125, 1304

\bibitem[{{Bica} {et~al.}(2008){Bica}, {Bonatto}, {Dutra}, \&
  {Santos}}]{betal08}
{Bica}, E., {Bonatto}, C., {Dutra}, C.~M., \& {Santos}, J.~F.~C. 2008, \mnras,
  389, 678

\bibitem[{{Bressan} {et~al.}(2012){Bressan}, {Marigo}, {Girardi}, {Salasnich},
  {Dal Cero}, {Rubele}, \& {Nanni}}]{betal12}
{Bressan}, A., {Marigo}, P., {Girardi}, L., {et~al.} 2012, \mnras, 427, 127

\bibitem[{{Cardelli} {et~al.}(1989){Cardelli}, {Clayton}, \&
  {Mathis}}]{cetal89}
{Cardelli}, J.~A., {Clayton}, G.~C., \& {Mathis}, J.~S. 1989, \apj, 345, 245

\bibitem[{{Cioni} {et~al.}(2011){Cioni}, {Clementini}, {Girardi}, {Guandalini},
  {Gullieuszik}, {Miszalski}, {Moretti}, {Ripepi}, {Rubele}, {Bagheri},
  {Bekki}, {Cross}, {de Blok}, {de Grijs}, {Emerson}, {Evans}, {Gibson},
  {Gonzales-Solares}, {Groenewegen}, {Irwin}, {Ivanov}, {Lewis}, {Marconi},
  {Marquette}, {Mastropietro}, {Moore}, {Napiwotzki}, {Naylor}, {Oliveira},
  {Read}, {Sutorius}, {van Loon}, {Wilkinson}, \& {Wood}}]{cetal11}
{Cioni}, M.-R.~L., {Clementini}, G., {Girardi}, L., {et~al.} 2011, \aap, 527,
  A116

\bibitem[{{de Grijs} \& {Goodwin}(2008)}]{dgetal08}
{de Grijs}, R. \& {Goodwin}, S.~P. 2008, \mnras, 383, 1000

\bibitem[{{de Grijs} {et~al.}(2014){de Grijs}, {Wicker}, \& {Bono}}]{dgetal14}
{de Grijs}, R., {Wicker}, J.~E., \& {Bono}, G. 2014, \aj, 147, 122

\bibitem[{{Dieball} \& {Grebel}(2000)}]{dg2000}
{Dieball}, A. \& {Grebel}, E.~K. 2000, \aap, 358, 897

\bibitem[{{Geha} {et~al.}(1998){Geha}, {Holtzman}, {Mould}, {Gallagher},
  {Watson}, {Cole}, {Grillmair}, {Stapelfeldt}, {Ballester}, {Burrows},
  {Clarke}, {Crisp}, {Evans}, {Griffiths}, {Hester}, {Scowen}, {Trauger}, \&
  {Westphal}}]{getal98}
{Geha}, M.~C., {Holtzman}, J.~A., {Mould}, J.~R., {et~al.} 1998, \aj, 115, 1045

\bibitem[{{Geisler}(1996)}]{g96}
{Geisler}, D. 1996, \aj, 111, 480

\bibitem[{{Glatt} {et~al.}(2011){Glatt}, {Grebel}, {Jordi}, {Gallagher}, {Da
  Costa}, {Clementini}, {Tosi}, {Harbeck}, {Nota}, {Sabbi}, \&
  {Sirianni}}]{getal11}
{Glatt}, K., {Grebel}, E.~K., {Jordi}, K., {et~al.} 2011, \aj, 142, 36

\bibitem[{{Haschke} {et~al.}(2011){Haschke}, {Grebel}, \& {Duffau}}]{hetal11}
{Haschke}, R., {Grebel}, E.~K., \& {Duffau}, S. 2011, \aj, 141, 158

\bibitem[{{Lada} \& {Lada}(2003)}]{ll03}
{Lada}, C.~J. \& {Lada}, E.~A. 2003, \araa, 41, 57

\bibitem[{{Mackey} \& {Gilmore}(2004)}]{mg04}
{Mackey}, A.~D. \& {Gilmore}, G.~F. 2004, \mnras, 352, 153

\bibitem[{{Nayak} {et~al.}(2016){Nayak}, {Subramaniam}, {Choudhury}, {Indu}, \&
  {Sagar}}]{nayaketal2016}
{Nayak}, P.~K., {Subramaniam}, A., {Choudhury}, S., {Indu}, G., \& {Sagar}, R.
  2016, \mnras, 463, 1446

\bibitem[{{No{\"e}l} {et~al.}(2009){No{\"e}l}, {Aparicio}, {Gallart},
  {Hidalgo}, {Costa}, \& {M{\'e}ndez}}]{netal09}
{No{\"e}l}, N.~E.~D., {Aparicio}, A., {Gallart}, C., {et~al.} 2009, \apj, 705,
  1260

\bibitem[{{Olszewski} {et~al.}(1996){Olszewski}, {Suntzeff}, \&
  {Mateo}}]{oetal96}
{Olszewski}, E.~W., {Suntzeff}, N.~B., \& {Mateo}, M. 1996, \araa, 34, 511

\bibitem[{{Pandey} {et~al.}(2010){Pandey}, {Sandhu}, {Sagar}, \&
  {Battinelli}}]{pandeyetal2010}
{Pandey}, A.~K., {Sandhu}, T.~S., {Sagar}, R., \& {Battinelli}, P. 2010,
  \mnras, 403, 1491

\bibitem[{{Piatti}(2012)}]{p12a}
{Piatti}, A.~E. 2012, \mnras, 422, 1109

\bibitem[{{Piatti}(2014{\natexlab{a}})}]{p14c}
{Piatti}, A.~E. 2014{\natexlab{a}}, \mnras, 445, 2302

\bibitem[{{Piatti}(2014{\natexlab{b}})}]{p14}
{Piatti}, A.~E. 2014{\natexlab{b}}, \mnras, 440, 3091

\bibitem[{{Piatti}(2014{\natexlab{c}})}]{p14b}
{Piatti}, A.~E. 2014{\natexlab{c}}, \mnras, 437, 1646

\bibitem[{{Piatti}(2015)}]{p15}
{Piatti}, A.~E. 2015, \mnras, 451, 3219

\bibitem[{{Piatti}(2016)}]{p16}
{Piatti}, A.~E. 2016, \mnras [\eprint[arXiv]{1603.06803}]

\bibitem[{{Piatti}(2017)}]{p17}
{Piatti}, A.~E. 2017, \apjl, 834, L14

\bibitem[{{Piatti} \& {Bastian}(2016)}]{pb16a}
{Piatti}, A.~E. \& {Bastian}, N. 2016, \aap, 590, A50

\bibitem[{{Piatti} \& {Bica}(2012)}]{pb12}
{Piatti}, A.~E. \& {Bica}, E. 2012, \mnras, 425, 3085

\bibitem[{{Piatti} \& {Cole}(2017)}]{pc17}
{Piatti}, A.~E. \& {Cole}, A. 2017, \inpress

\bibitem[{{Piatti} {et~al.}(2017){Piatti}, {Cole}, \& {Emptage}}]{petal2017}
{Piatti}, A.~E., {Cole}, A., \& {Emptage}, B. 2017, \prepare

\bibitem[{{Piatti} {et~al.}(2015{\natexlab{a}}){Piatti}, {de Grijs}, {Ripepi},
  {Ivanov}, {Cioni}, {Marconi}, {Rubele}, {Bekki}, \& {For}}]{petal15b}
{Piatti}, A.~E., {de Grijs}, R., {Ripepi}, V., {et~al.} 2015{\natexlab{a}},
  \mnras, 454, 839

\bibitem[{{Piatti} {et~al.}(2015{\natexlab{b}}){Piatti}, {de Grijs}, {Rubele},
  {Cioni}, {Ripepi}, \& {Kerber}}]{petal15a}
{Piatti}, A.~E., {de Grijs}, R., {Rubele}, S., {et~al.} 2015{\natexlab{b}},
  \mnras, 450, 552

\bibitem[{{Piatti} \& {Geisler}(2013)}]{pg13}
{Piatti}, A.~E. \& {Geisler}, D. 2013, \aj, 145, 17

\bibitem[{{Piatti} {et~al.}(2012){Piatti}, {Geisler}, \& {Mateluna}}]{pietal12}
{Piatti}, A.~E., {Geisler}, D., \& {Mateluna}, R. 2012, \aj, 144, 100

\bibitem[{{Piatti} {et~al.}(2014){Piatti}, {Guandalini}, {Ivanov}, {Rubele},
  {Cioni}, {de Grijs}, {For}, {Clementini}, {Ripepi}, {Anders}, \&
  {Oliveira}}]{petal14b}
{Piatti}, A.~E., {Guandalini}, R., {Ivanov}, V.~D., {et~al.} 2014, \aap, 570,
  A74

\bibitem[{{Piatti} {et~al.}(2016){Piatti}, {Ivanov}, {Rubele}, {Marconi},
  {Ripepi}, {Cioni}, {Oliveira}, \& {Bekki}}]{petal2016}
{Piatti}, A.~E., {Ivanov}, V.~D., {Rubele}, S., {et~al.} 2016, \mnras, 460, 383

\bibitem[{{Popescu} {et~al.}(2012){Popescu}, {Hanson}, \&
  {Elmegreen}}]{poetal12}
{Popescu}, B., {Hanson}, M.~M., \& {Elmegreen}, B.~G. 2012, \apj, 751, 122

\bibitem[{{Rubele} {et~al.}(2012){Rubele}, {Kerber}, {Girardi}, {Cioni},
  {Marigo}, {Zaggia}, {Bekki}, {de Grijs}, {Emerson}, {Groenewegen},
  {Gullieuszik}, {Ivanov}, {Miszalski}, {Oliveira}, {Tatton}, \& {van
  Loon}}]{retal12}
{Rubele}, S., {Kerber}, L., {Girardi}, L., {et~al.} 2012, \aap, 537, A106

\bibitem[{{Sarajedini}(1998)}]{s98}
{Sarajedini}, A. 1998, \aj, 116, 738

\bibitem[{{Sharma} {et~al.}(2010){Sharma}, {Borissova}, {Kurtev}, {Ivanov}, \&
  {Geisler}}]{shetal10}
{Sharma}, S., {Borissova}, J., {Kurtev}, R., {Ivanov}, V.~D., \& {Geisler}, D.
  2010, \aj, 139, 878

\bibitem[{{Smecker-Hane} {et~al.}(2002){Smecker-Hane}, {Cole}, {Gallagher}, \&
  {Stetson}}]{smeckeretal2002}
{Smecker-Hane}, T.~A., {Cole}, A.~A., {Gallagher}, III, J.~S., \& {Stetson},
  P.~B. 2002, \apj, 572, 1083

\bibitem[{{Stetson} {et~al.}(1990){Stetson}, {Davis}, \& {Crabtree}}]{setal90}
{Stetson}, P.~B., {Davis}, L.~E., \& {Crabtree}, D.~R. 1990, in Astronomical
  Society of the Pacific Conference Series, Vol.~8, CCDs in astronomy, ed.
  G.~H. {Jacoby}, 289--304

\bibitem[{{Subramanian} \& {Subramaniam}(2009)}]{ss09}
{Subramanian}, S. \& {Subramaniam}, A. 2009, \aap, 496, 399

\bibitem[{{Udalski}(2003)}]{u03}
{Udalski}, A. 2003, \actaa, 53, 291

\bibitem[{{Zaritsky} {et~al.}(2004){Zaritsky}, {Harris}, {Thompson}, \&
  {Grebel}}]{zetal04}
{Zaritsky}, D., {Harris}, J., {Thompson}, I.~B., \& {Grebel}, E.~K. 2004, \aj,
  128, 1606

\end{thebibliography}


\begin{appendix} 
\section{OGLE-CL\,LMC\,377 images}

\begin{figure*}
	\includegraphics[width=\textwidth]{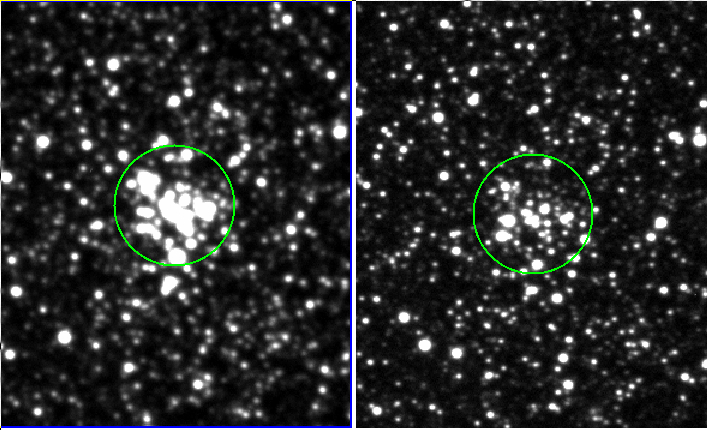}
    \caption{$C$ (left) and $R$ (right) images centred on OGLE-CL\,LMC\,377. The
circles are as in Fig.~\ref{fig:fig2}.}
\end{figure*}

\end{appendix}

\end{document}